\documentclass[aps,prb,reprint,twocolumn,showpacs,floatfix,superscriptaddress,nofootinbib]{revtex4}
\usepackage{graphicx}
\usepackage{amssymb}
\usepackage{amsmath}
\usepackage{lmodern}
\usepackage{color}
\usepackage{hyperref}
\usepackage{empheq}
\usepackage{epstopdf}
\usepackage{amsmath}

\begin{document}

\title{Modulated harmonic wave in series connected discrete Josephson transmission line: the discrete calculus approach}

\author{Eugene Kogan}
\email{Eugene.Kogan@biu.ac.il}
\affiliation{Department of Physics, Bar-Ilan University, Ramat-Gan 52900, Israel}
\affiliation{Max-Planck-Institut fur Physik komplexer Systeme
Dresden 01187, Germany}

\begin{abstract}
We consider the modulated harmonic wave in the discrete series connected Josephson transmission line (JTL). We formulate the approach to the modulation problems for
discrete wave equations based on discrete calculus. We check up the approach by applying it to the Fermi-Pasta-Ulam-Tsingou type problem. Applying the approach to the discrete JTL,   we obtain the equation describing the modulation amplitude, which
turns out to be the defocusing   nonlinear Schr\"odinger (NLS) equation. We compare the profile of the single soliton solution of the NLS with that of the soliton obtained in our previous publication.

\end{abstract}

\date{\today}

\maketitle


\section{Introduction}

In our previous publication \cite{kogan} we considered propagation of kinks, solitons and shocks  along the Josephson transmission line (JTL).
In the present short note we want to consider a different kind of excitations
 -- modulated harmonic waves \cite{solitons}.
In literature,  JTLs have been extensively discussed
in connection with travelling wave amplifies. The main interest was in interaction of small number of harmonic waves (pump, signal and idler; see e.g. \cite{brien} and references there). An exception to this statement is Ref. \cite{grimso}, where the wave packets are considered (implicitly) in the analysis, but single-frequency excitations are used for visualizing results. Hence the topic of modulated harmonic waves in JTLs is not very well explored.
In the present paper we will  show that the modulation amplitude is described by the defocusing nonlinear Schr\"odinger (NLS) equation \cite{kosevich,kivshar,kevrekidis}.

The rest of the paper is constructed as follows. In Section
\ref{discr} we write down equations describing JTL and present the definition of the modulated harmonic wave.
In Section \ref{linear} we present our approach based on discrete calculus (DC) and show that in  linear approximation,
 equation describing the modulation amplitude turns out to be linear Schr\"odinger equation.
Nonlinear problem is considered in  Section \ref{nls} in the framework of the DC approach, and the equation
describing the modulation amplitude turns out to be the defocusing nonlinear Schr\"odinger equation.
In Section \ref{fpu} we show that the DC approach, being applied to Fermi-Pasta-Ulam-Tsingou type problem gives the results identical to the known ones.
In Section \ref{ds9} we compare dark solitons, known to exist for the defocusing nonlinear Schr\"odinger equation, with the solitons in the JTL,  obtained by us previously.
We conclude in Section \ref{conc}.
In the Appendix  we compare the results of the DC approach
with those obtained in the framework of the fourier integral representation of the solution.

\section{Discrete Josephson transmission line}
\label{discr}

\begin{figure}[h]
\includegraphics[width=\columnwidth]{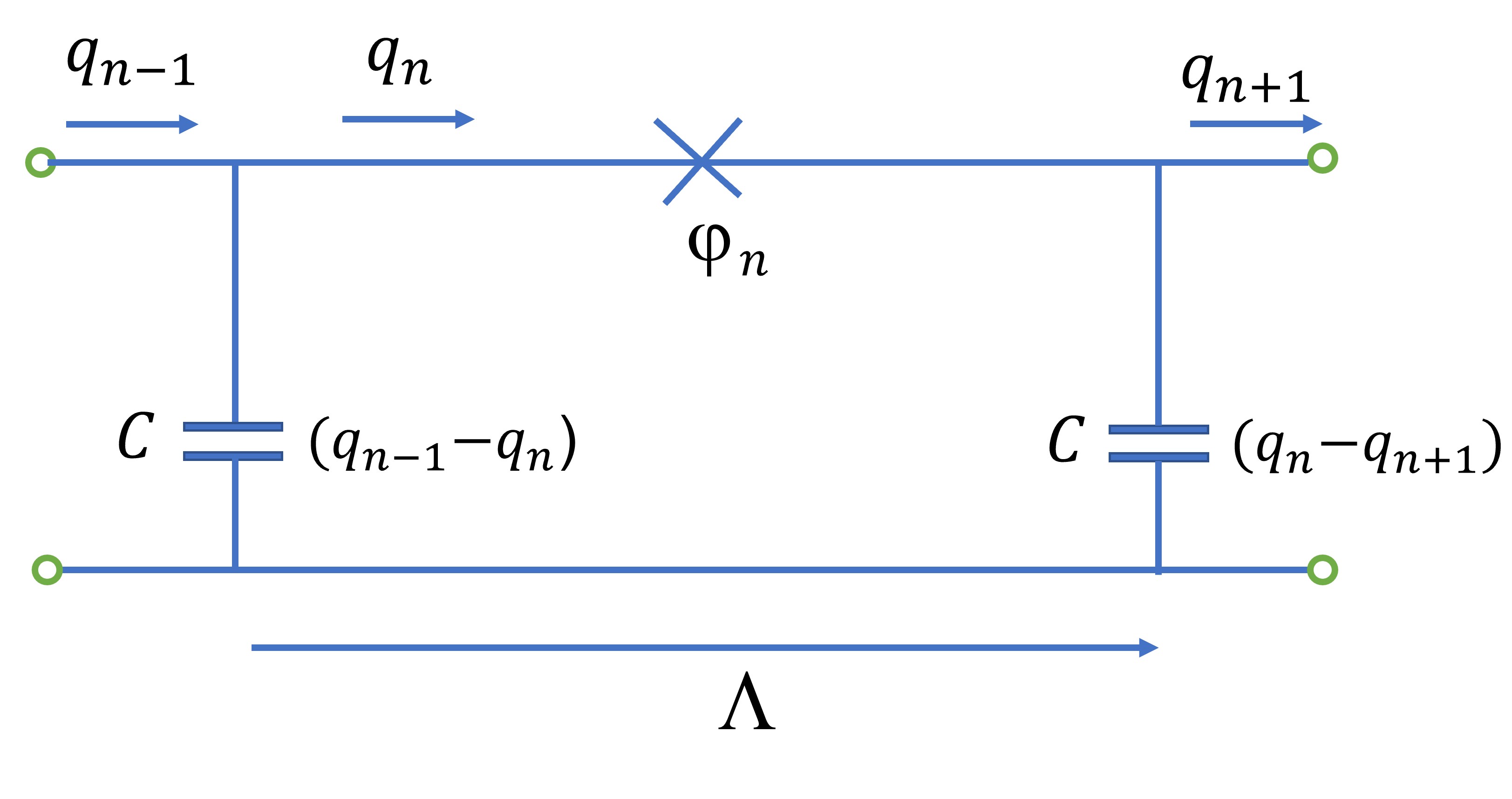}
\vskip -.5cm
\caption{Discrete  JTL. }
\label{trans1}
\end{figure}

Consider the  model of JTL constructed from identical Josephson junctions (JJ) and capacitors, which is shown in Figure \ref{trans1}.
We take
as dynamical variables  the phase differences (which we for brevity will call just phases) $\varphi_n$ across the  JJ
and the charges $q_n$ which have passed through the  JJ.
The  circuit equations are
\begin{subequations}
\label{ave7}
\begin{alignat}{4}
\frac{\hbar}{2e}\frac{d \varphi_n}{d t}&=\frac{1}{C}\left(q_{n+1}-2q_{n}+q_{n-1}\right)  \, ,\label{ave7a}\\
\frac{dq_n}{dt} &=   I_c\sin\varphi_n \, ,\label{ave7b}
\end{alignat}
\end{subequations}
where    $C$ is the capacitance, and  $I_c$ is the critical current of the JJ.
Differentiating Eq. (\ref{ave7a}) with respect to $t$ and substituting $dq_n/dt$  from Eq. (\ref{ave7b}), we obtain closed equation for $\varphi_n$ \cite{kogan}
\begin{eqnarray}
\label{com}
\frac{d^2 \varphi_n}{d \tau^2}=\sin\varphi_{n+1}-2\sin\varphi_{n}+\sin\varphi_{n-1}\,,
\end{eqnarray}
where we have introduced the dimensionless time $\tau=t/\sqrt{L_JC}$,
and $L_J=\hbar/(2eI_c)$.

In the linear approximation, Eq. (\ref{com}) takes the form
\begin{eqnarray}
\label{com2}
\frac{d^2 \varphi_n}{d \tau^2}=\varphi_{n+1}-2\varphi_{n}+\varphi_{n-1}\,.
\end{eqnarray}
The dispersion law for the eigenmodes of Eq. (\ref{com2}) can be easily found \cite{kittel}
\begin{eqnarray}
\label{dispe}
\omega(k)=2\left|\sin\left(\frac{k}{2}\right)\right|\,,\;\;k\in(-\pi,\pi]\,.
\end{eqnarray}

We will consider a modulated harmonic wave, that is a
carrier wave of  wave number $k_0$ (for the sake of definiteness we take $k_0>0$) and frequency $\omega_0=2\sin (k_0/2)$, modulated
by a waveform $a$, which varies slowly in time and
space compared to the variations of the carrier wave
\begin{eqnarray}
\label{pac}
\varphi(n,\tau)=\psi(n,\tau)+\psi^*(n,\tau)=a_nb_n+a_n^*b_n^*\,,
\end{eqnarray}
where
\begin{eqnarray}
\label{ab}
a_n=a(n,\tau)\,,\;\;\;\; b_n=e^{i(k_0n-\omega_0\tau)}\,,
\end{eqnarray}

\section{Linear approximation}
\label{linear}

Let us start from analysing  Eq. (\ref{com2}).
Because the equation is linear, we  obtain decoupled  equations for $\psi$ and $\psi^*$, and equation for $\psi$ becomes
\begin{eqnarray}
\label{com2b}
\frac{d^2 \psi_n}{d \tau^2}=\psi_{n+1}-2\psi_{n}+\psi_{n-1}\,.
\end{eqnarray}
Substituting $\psi_n=a_nb_n$
into (\ref{com2b}) we obtain
\begin{eqnarray}
\label{cm4}
a_n\frac{d^2 b_n}{d \tau^2}+2\frac{d a_n}{d \tau}\frac{d b_n}{d \tau}+\frac{d^2 a_n}{d \tau^2}b_n\nonumber\\
=a_{n+1}b_{n+1}-2a_nb_n+a_{n-1}b_{n-1}\,.
\end{eqnarray}
We first focus on the r.h.s. of (\ref{cm4})  and
use the elementary formula  for the discrete second derivative of the product of two functions
\begin{eqnarray}
\label{com3}
a_{n+1}b_{n+1}-2a_nb_n+a_{n-1}b_{n-1}\nonumber\\
=a_n\left(b_{n+1}-2b_n+b_{n-1}\right)+\left(a_{n+1}-a_n\right)\left(b_{n+1}
-b_n\right)\nonumber\\
+\left(a_{n}-a_{n-1}\right)\left(b_n
-b_{n-1}\right)
+\left(a_{n+1}-2a_n+a_{n-1}\right)b_n\,.
\end{eqnarray}
 Then the first term of the r.h.s. of (\ref{com3}) cancels against the $a_nd^2b_n/d\tau^2$-term on the l.h.s. of (\ref{cm4}), because $b_n$ is by itself the solution of (\ref{com2}),  and we  arrive to the equation
\begin{eqnarray}
\label{com4}
2\frac{d a_n}{d \tau}\frac{d b_n}{d \tau}+\frac{d^2 a_n}{d \tau^2}b_n
=\left(a_{n+1}-a_n\right)\left(b_{n+1}
-b_n\right)\nonumber\\
+\left(a_{n}-a_{n-1}\right)\left(b_n-b_{n-1}\right)
+\left(a_{n+1}-2a_n+a_{n-1}\right)b_n,\nonumber\\
\end{eqnarray}
or, more explicitly (taking into account (\ref{dispe})),
\begin{eqnarray}
\label{com444}
-4i\sin\left(\frac{k_0}{2}\right)\frac{d a_n}{d \tau}&+&\frac{d^2 a_n}{d \tau^2}
=i\sin k_0\left(a_{n+1}-a_{n-1}\right)\nonumber\\
&+&\cos k_0\left(a_{n+1}-2a_n+a_{n-1}\right).
\end{eqnarray}
In the lowest order (with respect to the ratio of the effective modulation wave vector and $k_0$) approximation we should equate the first terms in the l.h.s. and r.h.s.
of (\ref{com444}) thus obtaining
\begin{eqnarray}
\label{om444}
\frac{d a_n}{d \tau}
=-\frac{1}{2}\cos\left(\frac{k_0}{2}\right)\left(a_{n+1}-a_{n-1}\right)
\end{eqnarray}
and, hence,
\begin{eqnarray}
\label{om454}
\frac{d^2 a_n}{d \tau^2}
=\frac{1}{4}\cos^2\left(\frac{k_0}{2}\right)\left(a_{n+2}-2a_n+a_{n-2}\right)\,.
\end{eqnarray}
Substituting (\ref{om454}) into (\ref{com444}) we obtain a discrete linear Schr\"odinger equation (with an additional convection term that can be removed by a change of reference frame)
\begin{eqnarray}
\label{c4}
-4i\sin\left(\frac{k_0}{2}\right)\frac{d a_n}{d \tau}
=i\sin k_0\left(a_{n+1}-a_{n-1}\right)\nonumber\\
+\cos k_0\left(a_{n+1}-2a_n+a_{n-1}\right)\nonumber\\
-\frac{1}{4}\cos^2\left(\frac{k_0}{2}\right)\left(a_{n+2}-2a_n+a_{n-2}\right)\,.
\end{eqnarray}

Substituting continuous variable $x$ for the discrete variable $n$ as the argument of $a$ and expanding around $x=n$, we can present the quantities in the parentheses of (\ref{c4}) as
\begin{subequations}
\label{ae7}
\begin{alignat}{4}
a_{n+1}-a_{n-1}&=2\frac{\partial a}{\partial x}  \, ,\label{ae7a}\\
a_{n+1}-2a_n+a_{n-1} &=  \frac{\partial^2 a}{\partial x^2}  \, ,\label{ae7b}\,\\
a_{n+2}-2a_n+a_{n-2} &=  4\frac{\partial^2 a}{\partial x^2}\,,
\end{alignat}
\end{subequations}
and after simple algebra the equation takes the form
of the linear Schr\"odinger equation
\begin{eqnarray}
\label{lin}
i\left(\frac{\partial a}{\partial \tau}+v_g\frac{\partial a}{\partial x}\right)-D\frac{\partial ^2a}{\partial x^2}=0\,,
\end{eqnarray}
where  the group velocity $v_g$ and the  coefficient $D$ are
\begin{eqnarray}
v_g=\cos\left(\frac{k_0}{2}\right)\,,\;\;\;\;\; D=\frac{1}{4}\sin\left(\frac{k_0}{2}\right)\,.
\end{eqnarray}

When one looks at (\ref{lin}) a natural question appears: If the approximation of the underlying chain of coupled junctions by the quasi-continuum limit is extended, what will be the next-order terms in the resulting  equation? We can answer this question by inspection of (\ref{ae7}). In case of the extension mentioned above,  in the derived linear Schr\"odinger equation there would appear additional (quasi)
drift term with the third derivative with respect to $x$, and additional  term with the forth derivative with respect to $x$.

\section{The nonlinear Schr\"odinger equation}
\label{nls}

Now let us return to Eq. (\ref{com}). Presenting $\varphi$ as  in (\ref{pac}) and expanding the sine in series we get
\begin{eqnarray}
\sin\varphi=\sum_{m=0}^{\infty}\frac{(-1)^m}{(2m+1)!}(\psi+\psi^*)^{2m+1}\,.
\end{eqnarray}
Keeping only the first harmonics (the rotating wave approximation), we will present $\sin\varphi$ as
\begin{eqnarray}
\sin\varphi&=&
\left(\psi+\psi^*\right)\sum_{m=0}^{\infty}\frac{(-1)^m}{m!(m+1)!}|\psi|^{2m}\nonumber\\
&=&\left(\psi+\psi^*\right)\frac{J_1(2|\psi|)}{|\psi|}\,,
\end{eqnarray}
where $J_1$ is the Bessel function.
Thus we again obtain decoupled  equations for $\psi$ and $\psi^*$, and equation for $\psi$ becomes
(compare with (\ref{com2b}))
\begin{eqnarray}
\label{om2}
\frac{d^2 \psi_n}{d \tau^2}=\frac{J_1(2|\psi_{n+1}|)}{|\psi_{n+1}|}\psi_{n+1}
-2\frac{J_1(2|\psi_{n}|)}{|\psi_{n}|}\psi_{n}\nonumber\\
+\frac{J_1(2|\psi_{n-1}|)}{|\psi_{n-1}|}\psi_{n-1}\,.
\end{eqnarray}
Note that in the lowest nontrivial order (\ref{om2}) takes the form
\begin{eqnarray}
\label{om2b}
\frac{d^2 \psi_n}{d \tau^2}=\left(1-\frac{|\psi_{n+1}|^2}{2}\right)\psi_{n+1}
-2\left(1-\frac{|\psi_{n}|^2}{2}\right)\psi_{n}\nonumber\\
+\left(1-\frac{|\psi_{n-1}|^2}{2}\right)\psi_{n-1}\,.
\end{eqnarray}
Using (\ref{pac}) we obtain
\begin{eqnarray}
\label{com44}
&&2\frac{d a_n}{d \tau}\frac{d b_n}{d \tau}+\frac{d^2 a_n}{d \tau^2}b_n
=\left(a_{n+1}-a_n\right)\left(b_{n+1}
-b_n\right)\nonumber\\
&&+\left(a_{n}-a_{n-1}\right)\left(b_n
-b_{n-1}\right)
+\left(a_{n+1}-2a_n+a_{n-1}\right)b_n\nonumber\\
&&-d.s.d.\,,
\end{eqnarray}
where the last term in the r.h.s. of (\ref{com44}) is
the discrete second derivative (d.s.d.) of the product of three quantities
\begin{eqnarray}
\label{ds}
d.s.d.=a_{n+1}g_{n+1}b_{n+1}-2a_{n}g_nb_{n}
+a_{n-1}g_{n-1}b_{n-1},
\end{eqnarray}
and
\begin{eqnarray}
g_n\equiv g\left(|a_{n}|^2\right)=1-\frac{J_1(2|a_n|)}{|a_n|}\,.
\end{eqnarray}

The first three terms in the r.h.s. of (\ref{com44}) are identical to the r.h.s. of (\ref{com4}). The forth term can be treated in a simpler way. In fact, the difficulty of treating the r.h.s. of  (\ref{com4}) was connected with the fact that 
in the expansion 
with respect to the ratio of the effective modulation wave vector and $k_0$,
the lowest order   term was canceled with the appropriate term in the l.h.s.
Thus we have to take into account the next order terms. There is no such cancellation for the d.s.d. term in the r.h.s. of (\ref{com44}), so the term can be considered in the lowest order approximation. 
Among the  three quantities $a_n$, $g_n$ and $b_n$, two ($a_n$ and $g_n$) change slowly with $n$, and the third quantity ($b_n$) changes fast. This is why in the r.h.s. of (\ref{ds})  we can ignore the difference between $a_n$, $a_{n-1}$ and $a_{n+1}$ (and between $g_n$, $g_{n-1}$ and $g_{n+1}$) and present the equation as
\begin{eqnarray}
\label{com3c}
d.s.d.=a_{n}g_{n}\left(b_{n+1}-2b_{n}+b_{n-1}\right)\,.
\end{eqnarray}
Proceeding as in Section \ref{linear},
 we obtain, instead of (\ref{lin}), the defocusing NLS \cite{karpman,kosevich,kivshar,kevrekidis}
\begin{eqnarray}
\label{sch0}
i\left(\frac{\partial a}{\partial \tau}+v_g\frac{\partial a}{\partial x}\right)-D\frac{\partial ^2a}{\partial x^2}
+4Dg\left(|a|^2\right)a=0\,.
\end{eqnarray}
In the lowest nontrivial order with respect to $|a|^2$, Eq. (\ref{sch0}) is reduced to
\begin{eqnarray}
\label{linda}
i\left(\frac{\partial a}{\partial \tau}+v_g\frac{\partial a}{\partial x}\right)-D\frac{\partial ^2a}{\partial x^2}
+2D|a|^2a=0\,.
\end{eqnarray}
As an additional support for the validity of Eq. (\ref{linda}) we will "rederive" it in the Appendix.

\section{The Fermi-Pasta-Ulam-Tsingou problem}
\label{fpu}

To check up our DC method, it would be appropriate to apply it to the Fermi-Pasta-Ulam-Tsingou (FPUT) problem \cite{gallavotti}. The FPUT analog of (\ref{com}) would be \cite{james,james2}
\begin{eqnarray}
\label{combocho}
\frac{d^2 \varphi_n}{d \tau^2}=\sin\left(\varphi_{n+1}-\varphi_n\right)+\sin\left(\varphi_{n-1}-\varphi_n\right)\,.
\end{eqnarray}
In the rotating wave approximation we  again have the decoupling of the equations for $\psi$ and $\psi^*$. Expanding sine in Taylor series and keeping the two lowest order nonzero terms, we obtain equation for $\psi$ in the form
\begin{eqnarray}
\label{combocho2}
\frac{d^2 \psi_n}{d \tau^2}=\psi_{n+1}-\psi_n+\psi_{n-1}-\psi_n\nonumber\\
-\frac{1}{2}\left[\left|\psi_{n+1}^2\right|\psi_{n+1}-\psi_{n+1}^2\psi_n^*
-2\left|\psi_{n}^2\right|\psi_n\right.\nonumber\\
+\psi_{n}^2\psi_{n+1}^*-2\left|\psi_{n+1}^2\right|\psi_{n}
+2\left|\psi_{n}^2\right|\psi_{n+1}\nonumber\\
+\left|\psi_{n-1}^2\right|\psi_{n-1}-\psi_{n-1}^2\psi_n^*\nonumber\\
\left.+\psi_{n}^2\psi_{n-1}^*-2\left|\psi_{n-1}^2\right|\psi_{n}
+2\left|\psi_{n}^2\right|\psi_{n-1}\right]\,.
\end{eqnarray}
Following the example of Section \ref{nls} we again obtain the defocusing NLS
\begin{eqnarray}
\label{linda6}
i\left(\frac{\partial a}{\partial \tau}+v_g\frac{\partial a}{\partial x}\right)
-D\frac{\partial ^2a}{\partial x^2}+D'|a|^2a=0\,,
\end{eqnarray}
only this time
\begin{eqnarray}
D'=\frac{1}{8b_n\sin\left(\frac{k_0}{2}\right)}\left(3b_{n+1}+3b_{n-1}-b_{n+1}^2b_n^*
\right.\nonumber\\
\left.-b_{n-1}^2b_n^*-6b_n+b_{n}^2b_{n+1}^*+b_{n}^2b_{n-1}^*\right)\,.
\end{eqnarray}
After simple algebra we obtain
\begin{eqnarray}
D'=2\sin^3\left(\frac{k_0}{2}\right)\,.
\end{eqnarray}
Equation (\ref{linda6}) exactly coincides with the result of Ref. \cite{fly}, in the appropriate  particular case. (In that  Reference, equation  more general than (\ref{combocho}) is considered.)

\section{Dark solitons}
\label{ds9}

The defocusing NLS equation  has an interesting type of solutions,
called dark solitons \cite{zakharov,veks,nateo,triki}. Using the opportunity, we would like to  present here the derivation of  these  solutions, borrowed to some extent from the book \cite{kosevich} (which, to the best of our knowledge, was never translated into English).

Making the transformation
\begin{eqnarray}
\xi=\left(x-v_g\tau\right),\;\;\;\tau'=D\tau,
\end{eqnarray}
we can present (\ref{linda}) as
\begin{eqnarray}
\label{sch}
i\frac{\partial a}{\partial \tau'}-\frac{\partial ^2a}{\partial \xi^2}+2|a|^2a=0\,.
\end{eqnarray}

\begin{figure}[h]
\includegraphics[width=.7\columnwidth]{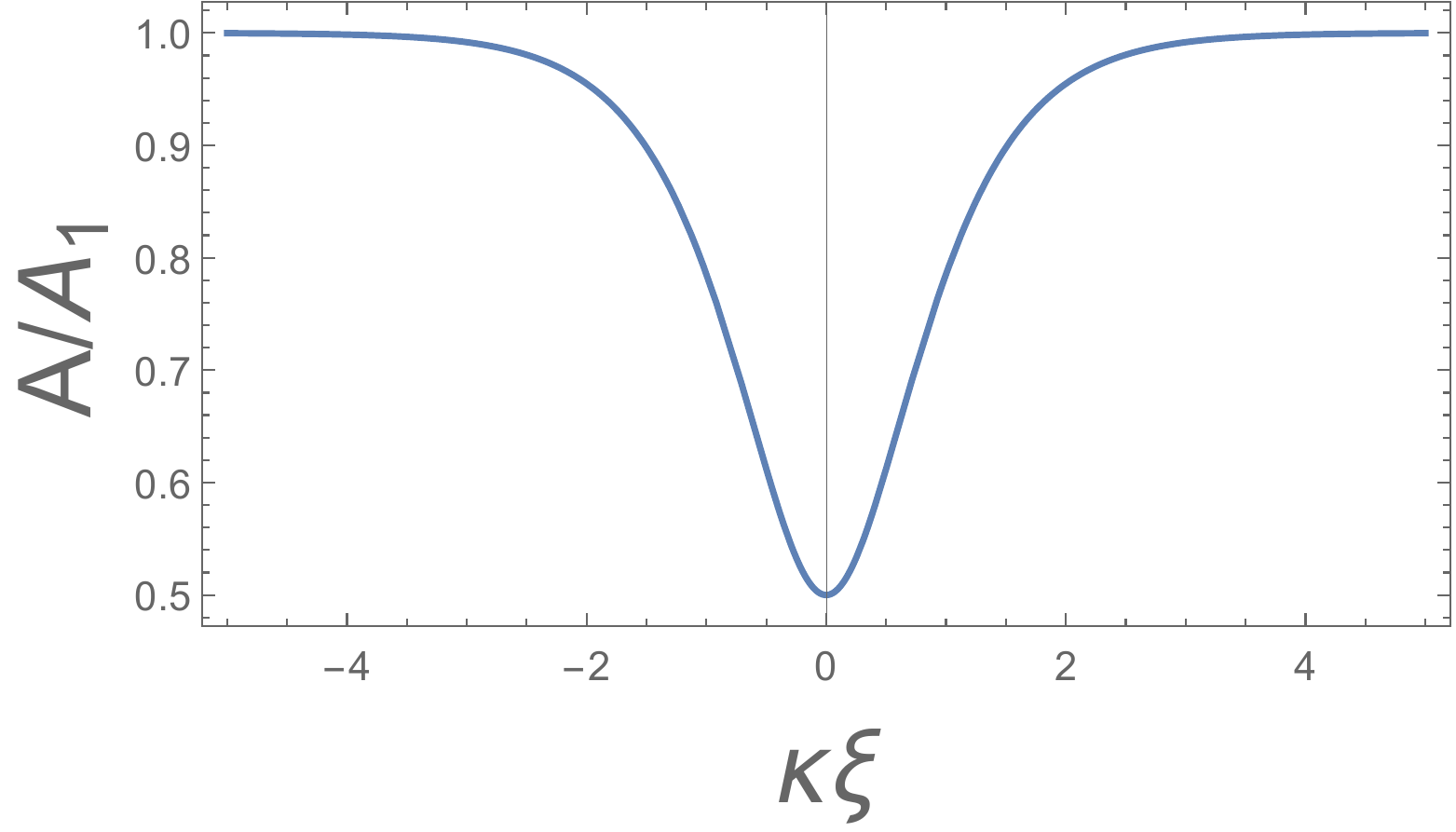}
\caption{Dark soliton as given by Eq. (\ref{rho}. We have chosen $A_0=.5A_1$.}
 \label{3}
\end{figure}

Looking for the   solutions of (\ref{sch}) in the form
\begin{eqnarray}
\label{127}
a=Ae^{i\Phi}e^{i\Omega \tau'},
\end{eqnarray}
where $\Omega$ is a constant, and the amplitude $A$ and the phase $\Phi$ of $a$ are some functions of $\xi$,
we obtain for those functions the system of equations
\begin{subequations}
\label{si}
\begin{alignat}{4}
\frac{d }{d \xi}\left(A^2\frac{d\Phi}{d\xi}\right)&=0\, ,\label{103}\\
\frac{d^2A}{d\xi^2}-A\left(\frac{d\Phi}{d\xi}\right)^2-2A^3+\Omega A&= 0 \, .\label{104}
\end{alignat}
\end{subequations}
From (\ref{103}) we obtain
\begin{eqnarray}
\frac{d\Phi}{d\xi}=\frac{c}{A^2}\,,
\end{eqnarray}
where $c$ is the integration constant.
Substituting this expression into (\ref{104})
we
obtain
\begin{eqnarray}
\label{101}
\frac{d^2A}{d\xi^2}-\frac{c^2}{ A^3}-2A^3+\Omega A &= 0\,.
\end{eqnarray}
Integrating, we
obtain the differential equation of the first order
\begin{eqnarray}
\label{128}
\left(\frac{dA}{d\xi}\right)^2+\frac{c^2}{ A^2}-A^4+\Omega A^2=E\,,
\end{eqnarray}
where $E$ is another integration constant. We are looking for the solutions satisfying the conditions
\begin{eqnarray}
\lim_{\xi\to\pm\infty}A=A_1.
\end{eqnarray}
Expressing the constants of integration through $\Omega$ and $A_1$
\begin{eqnarray}
c^2=A_1^4\left(\Omega-2A_1^2\right)\,,\hskip 1 cm E=A_1^2\left(2\Omega-3A_1^2\right)\,,
\end{eqnarray}
we may present (\ref{128}) as
\begin{eqnarray}
\label{129}
\left(\frac{d\rho}{d\xi}\right)^2=4\left(\rho-\rho_1\right)^2\left(\rho+2\rho_1-\Omega\right),
\end{eqnarray}
where we have introduced $\rho=A^2$. Equation (\ref{129}) can be easily integrated, and we obtain
\begin{eqnarray}
\label{rho}
\rho=\rho_1-\frac{\kappa^2}{\cosh^2\kappa \xi},
\end{eqnarray}
where $\kappa^2=3\rho_1-\Omega$. It is convenient to present Eq. $\kappa$ as
\begin{eqnarray}
\label{rho2}
\kappa^2=A_1^2-A_0^2,
\end{eqnarray}
where $A_0$ is the minimal value of $A$ (achieved at $\xi=0$).
The function $A(\xi)$ defined by Eq. (\ref{rho})
is presented in Figure \ref{3}.

Using immobile soliton solution and the property of Galilean invariance of (\ref{sch}), it is easy to obtain a solution describing moving soliton. In general, if $a(\xi,\tau')$ is a solution of (\ref{sch}), so is $e^{i(2V\xi+V^2\tau')}a(\xi-V\tau',\tau')$ for arbitrary $V$.

It would be appropriate to compare the soliton given by Eq. (\ref{rho}) with  the solitons in the JTL presented in our previous publication \cite{kogan}.  Those solitons were characterized by the Josephson phase,
asymptotically constant at both ends of the line
\begin{eqnarray}
\lim_{x\to\pm\infty}\varphi=\varphi_1.
\end{eqnarray}
For the solitons described by (\ref{rho}), the phase  at both sides of the transmission line asymptotically coincides with that of a (high frequency) harmonic wave. In general, while carrier wave is all important in the present paper, there was no such wave whatsoever in our previous publication \cite{kogan}.

\begin{figure}[h]
\includegraphics[width=.7\columnwidth]{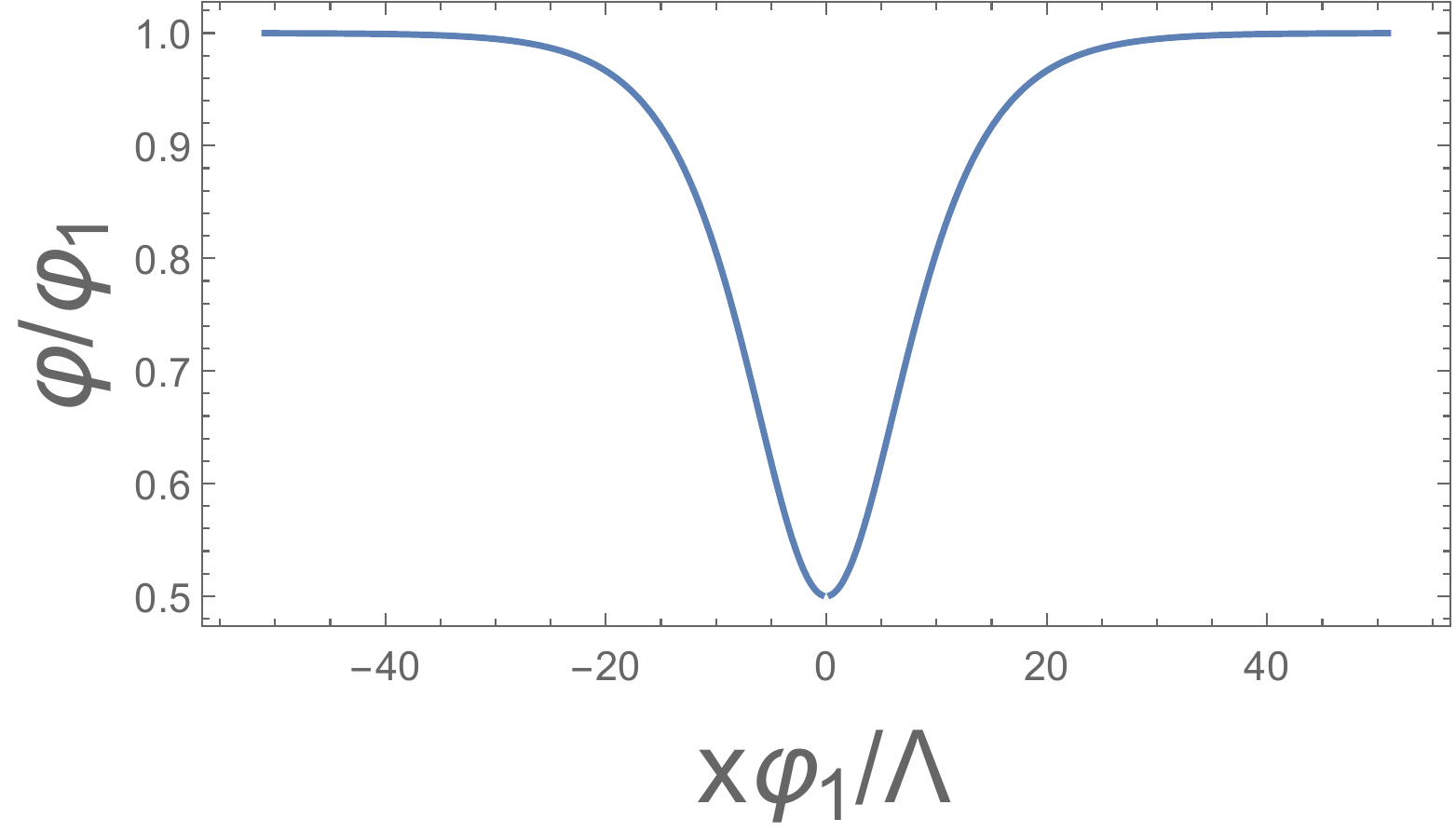}
\caption{The   soliton profile calculated in Ref. \cite{kogan}. ($\Lambda$ is the period of the JTL, which in the present publication was chosen to be equal to one.) }
 \label{trans3}
\end{figure}

However, 
it is worth to compare Figure \ref{3} with
Figure \ref{trans3} borrowed from our previous publication, and showing the soliton profile calculated there. Though we didn't use there the term dark soliton previously \cite{kogan}, we see that the curves are very much  similar: in  both cases the amplitude goes to a constant value when $x$ ($\xi$) goes to infinities and decreases in between.

\section{Conclusions}
\label{conc}

We consider the modulated harmonic wave in the discrete series connected Josephson transmission line (JTL). We formulate the approach to the modulation problems for
discrete wave equations based on discrete calculus. We check up the approach by applying it to the Fermi-Pasta-Ulam-Tsingou type problem. Applying the approach to the discrete JTL,   we obtain the equation describing the modulation amplitude, which
turns out to be the defocusing   nonlinear Schr\"odinger (NLS) equation. The NLS, being the normal form for
envelope waves, represents a universal model at the root of an extremely wide range of
physical and other natural phenomena and applications. Furthermore, due to its rich and
complex phenomenology, the NLS is also a paradigm for nonlinear spatio-temporal dynamics
and is at the forefront of intense and challenging mathematical research \cite{kevrekidis}.
We presented a new derivation of the single soliton solution of the NLS and compared its profiles with the profile of the soliton, obtained in our previous publication.

\begin{acknowledgments}

We are  grateful to  M. Goldstein, G. James and  B. Malomed  for their  insightful  comments. We are also very grateful to the anonymous Referees.

The work on the problem was initiated by my participation in the workshop  Coherent Structures: Current Developments and Future Challenges 4-8 July @ Oort. I would like to express my gratitude to the Lorentz Center for the hospitality and for the stimulating atmosphere.

\end{acknowledgments}

\begin{appendix}

\section{The  approach based on Fourier integral}

To additionally check up the discrete calculus approach  we present here an alternative derivation of some of the results obtained in the main body of the paper, based on Fourier integral representation of the solution. 

First consider the linear case.
Equation (\ref{lin}) can be put in a more general context  \cite{solitons}.
Let us present the solution $\psi(n,\tau)$ of a general linear equation  as the Fourier integral
\begin{eqnarray}
\label{psi}
\psi(n,\tau)=\int dk\psi_k\exp\left\{ikn-i\omega(k)\tau\right\}\,.
\end{eqnarray}
From (\ref{psi}) follows
\begin{eqnarray}
\label{aaaa}
a(x,\tau)=\int dk\psi_k\exp\left\{i(k-k_0)x-i\left[\omega(k)-\omega_0\right]\tau\right\}\,.\nonumber\\
\end{eqnarray}
Differentiating with respect to $x$ we obtain
\begin{eqnarray}
\label{aaa5}
i\frac{\partial a}{\partial x}=-\int dk(k-k_0)\psi_k\nonumber\\
\cdot\exp\left\{i(k-k_0)x-i\left[\omega(k)-\omega_0\right]\tau\right\}.
\end{eqnarray}
Differentiating with respect to $\tau$ we obtain
\begin{eqnarray}
\label{aaa}
i\frac{\partial a}{\partial\tau}=\int dk\left[\omega(k)-\omega_0\right]\psi_k\nonumber\\
\cdot\exp\left\{i(k-k_0)x-i\left[\omega(k)-\omega_0)\right]\tau\right\} .
\end{eqnarray}
Expanding  $\omega(k)$ with respect to $k$ up to the second order we get
\begin{eqnarray}
\label{8.14}
\omega(k)-\omega_0=\left(\frac{d\omega(k)}{dk}\right)_0(k-k_0)\nonumber\\
+\frac{1}{2}\left(\frac{d^2\omega(k)}{dk^2}\right)_0(k-k_0)^2
\end{eqnarray}
(the derivatives are calculated at $k=k_0$).
Comparing (\ref{aaa}) with (\ref{aaa5}) we obtain the equation for the modulation amplitude
\begin{eqnarray}
\label{linb}
i\left[\frac{\partial a}{\partial \tau}+\left(\frac{d\omega(k)}{dk}\right)_0\frac{\partial a}{\partial x}\right]+\frac{1}{2}\left(\frac{d^2\omega(k)}{dk^2}\right)_0\frac{\partial ^2a}{\partial x^2}=0\,.
\end{eqnarray}
For the dispersion law (\ref{dispe}), Eq. (\ref{linb}) exactly coincides with (\ref{lin}).

The mnemonic rule for obtaining the equation for amplitude,
modulating the harmonic wave $\exp\left\{ik_0n-i\omega(k_0)\tau\right\}$,  can be formulated as follows. Take the expansion of $\omega(k)$ with respect to $k$ up to the second order (\ref{8.14}),
replace $k-k_0$ by a spacial operator $-i\partial/\partial x$, and $\omega-\omega_0$ by a temporal operator $i\partial/\partial \tau$,
and let (\ref{8.14}) operate on the complex amplitude
function $a$.

If we  present the complex amplitude $a$ as
\begin{eqnarray}
a=|a|e^{i\theta}\,,
\end{eqnarray}
the complex equation (\ref{linb}) can be presented as two real equations
\begin{subequations}
\label{re}
\begin{alignat}{4}
\frac{\partial |a|^2}{\partial \tau}+\left(\frac{d\omega(k)}{dk}\right)_0\frac{\partial |a|^2}{\partial x}\nonumber\\
+\left(\frac{d^2\omega(k)}{dk^2}\right)_0\frac{\partial }{\partial x}
\left(|a|^2\frac{\partial \theta}{\partial x}\right)=0 \, ,
\label{rea}\\
\frac{\partial \theta}{\partial \tau}+\left(\frac{d\omega(k)}{dk}\right)_0\frac{\partial \theta}{\partial x}\nonumber\\
+\frac{1}{2}\left(\frac{d^2\omega(k)}{dk^2}\right)_0\left[\left(\frac{\partial \theta}{\partial x}\right)^2
-\frac{\partial^2 |a|}{\partial x^2}\right]=   0 \, .\label{reb}
\end{alignat}
\end{subequations}

It is interesting to compare Eq. (\ref{re}) with the equations borrowed from geometric optics \cite{whitham}. These equations describe a slowly
varying wavetrain by equations determining the propagation
of wave number and frequency
\begin{subequations}
\label{geome}
\begin{alignat}{4}
\frac{\partial |a|^2}{\partial \tau}+\frac{d\omega(k)}{dk}\frac{\partial |a|^2}{\partial x}
+\frac{d^2\omega(k)}{dk^2}|a|^2\frac{\partial k}{\partial x}&=  0\,,\label{geomea}\\
\frac{\partial k}{\partial \tau}+\frac{d\omega(k)}{dk}\frac{\partial k}{\partial x}&=  0\,.
\label{geomeb}
\end{alignat}
\end{subequations}
It is convenient to write down Eq. (\ref{geome}) in a more explicit form
by substituting $k=k_0+\partial\theta/\partial x$
\begin{subequations}
\label{geo}
\begin{alignat}{4}
\frac{\partial |a|^2}{\partial \tau}
&+\left(\frac{d\omega(k)}{dk}\right)_0\frac{\partial |a|^2}{\partial x}
+\left(\frac{d^2\omega(k)}{dk^2}\right)_0\frac{\partial \theta}{\partial x}\frac{\partial |a|^2}{\partial x} \nonumber\\
&+\left(\frac{d^2\omega(k)}{dk^2}\right)|a|^2\frac{\partial^2 \theta}{\partial x^2}=  0\,,\label{geoa}\\
\frac{\partial^2 \theta}{\partial x\partial\tau}&+\left(\frac{d\omega(k)}{dk}\right)_0
\frac{\partial^2 \theta}{\partial x^2}
+\left(\frac{d^2\omega(k)}{dk^2}\right)_0\frac{\partial \theta}{\partial x}
\frac{\partial^2 \theta}{\partial x^2}=  0\,.
\label{geob}
\end{alignat}
\end{subequations}

Equation (\ref{geoa}) is identical to (\ref{rea}).
To compare (\ref{geob}) with (\ref{reb}), let us differentiate the latter with respect to $x$.  We obtain
\begin{eqnarray}
\label{dif}
\frac{\partial \theta}{\partial x\partial \tau}
+\left(\frac{d\omega(k)}{dk}\right)_0\frac{\partial^2 \theta}{\partial x^2}
+\left(\frac{d^2\omega(k)}{dk^2}\right)_0\frac{\partial \theta}{\partial x}
\frac{\partial^2 \theta}{\partial x^2}\nonumber\\
-\frac{1}{2}\left(\frac{d^2\omega(k)}{dk^2}\right)_0
\frac{\partial^3 |a|}{\partial x^3}=   0\,.
\end{eqnarray}
 In the r.h.s. of Eq. (\ref{dif}) there is an additional term in comparison with (\ref{geob}), but this term is probably  negligible within the framework of the approximations made.

Now we can turn to the nonlinear case.
Equation (\ref{linda}) can be  "rederived" following the pattern
presented above.
The solution of (\ref{om2b}) with constant amplitude $a$ is
\begin{eqnarray}
\label{120}
\psi_n=ae^{i(kn-\omega t)}\,.
\end{eqnarray}
Substituting (\ref{120}) into (\ref{om2b}) we get the nonlinear dispersion law
\begin{eqnarray}
\label{aa}
\omega(k;|a|^2)=2\left(1-\frac{|a|^2}{4}\right)\left|\sin\left(\frac{k}{2}\right)\right|
=\omega(k)+\omega_2(k)|a|^2\, ,\nonumber\\
\end{eqnarray}
where $\omega_2=-|\sin (k/2)|/2$.
Hence (\ref{8.14}) should be modified to
\begin{eqnarray}
\label{8.14b}
\omega-\omega_0=\left(\frac{d\omega(k;0)}{dk}\right)_0(k-k_0)\nonumber\\
+\frac{1}{2}\left(\frac{d^2\omega(k;0)}{dk^2}\right)_0(k-k_0)^2
+\omega_2(k)|a|^2\,.
\end{eqnarray}

If we assume that in the nonlinear case, the formula (\ref{aaaa})  is modified to a self-consistent equation
\begin{eqnarray}
\label{aaab}
&&a(x,\tau)=\int dk\psi_k\nonumber\\
&\cdot&\exp\left\{i(k-k_0)x
-i\left[\omega(k;|a|^2)-\omega_0\right]\tau\right\}\,,
\end{eqnarray}
then (\ref{aaa5}) remains as it was in the linear case.
Differentiating (\ref{aaab}) with respect to $\tau$ we obtain
\begin{eqnarray}
\label{aab}
i\frac{\partial a}{\partial\tau}&=&\int dk\psi_k\exp\left\{i(k-k_0)x-i\left[\omega(k;|a|^2)-\omega_0\right]\tau\right\} \nonumber\\
&\cdot&\left[\left(\frac{d\omega(k)}{dk}\right)_0(k-k_0)
+\frac{1}{2}\left(\frac{d^2\omega(k)}{dk^2}\right)_0(k-k_0)^2\right.
\nonumber\\
&+&\left.\frac{\partial \omega(k;|a|^2)}{\partial |a|^2}|a|^2\right]\,.
\end{eqnarray}
Comparing (\ref{aab}) with (\ref{aaa5}) we get
\begin{eqnarray}
\label{linc}
i\left[\frac{\partial a}{\partial \tau}+\left(\frac{\partial\omega(k;a)}{\partial k}\right)_0\frac{\partial a}{\partial x}\right]&+&\frac{1}{2}\left(\frac{\partial^2\omega(k;a)}{\partial k^2}\right)_0\frac{\partial ^2a}{\partial x^2}\nonumber\\
-\frac{\partial \omega(k;|a|^2)}{\partial |a|^2}|a|^2&=&0\,,
\end{eqnarray}
which, for the dispersion law (\ref{aa}), exactly coincides with (\ref{linda}).
We hope that the identity of the results obtained by two different approximate methods gives us additional confidence in their validity.

\end{appendix}

\end{document}